\documentclass{elsart}
\usepackage{graphicx,color,epsfig}

\newcommand{\be}{\begin{equation}}
\newcommand{\ee}{\end{equation}}
\newcommand{\bn}{\begin{eqnarray}}
\newcommand{\en}{\end{eqnarray}}
\newcommand{\ba}{\begin{array}}
\newcommand{\ea}{\end{array}}
\newcommand{\bc}{\begin{center}}
\newcommand{\ec}{\end{center}}
\newcommand{\bml}{\begin{mathletters}}
\newcommand{\eml}{\end{mathletters}}

\begin{document}

\begin{frontmatter}
\title{Continuum Coupling and Single-Nucleon Overlap Integrals}
\author[a,b,c]{N. Michel,}
\author[a,b,d]{W. Nazarewicz,} and
\author[e]{M. P{\l}oszajczak}

\address[a]{Department of Physics and Astronomy, University of Tennessee, Knoxville, Tennessee 37996, U.S.A.}
\address[b]{Physics Division, Oak Ridge National Laboratory, Oak Ridge, Tennessee 37831, U.S.A.}
\address[c]{Department of Physics, Graduate School of Science, Kyoto University
                              Kitashirakawa, 606-8502, Kyoto, Japan}
\address[d]{Institute of Theoretical Physics, University of Warsaw, ul. Ho\.za 69, 00-681 Warsaw,
Poland }
\address[e]{Grand Acc\'el\'erateur National d'Ions Lourds (GANIL), CEA/DSM - CNRS/IN2P3,
BP 55027, F-14076 Caen Cedex, France}


\begin{abstract}
The presence of a particle continuum, both of a resonant and
non-resonant character, can significantly impact  spectroscopic
properties of weakly bound nuclei and excited nuclear states close  to,
and above,  the particle emission threshold. In the framework of the
continuum shell model in the complex momentum-plane,  the so-called
Gamow Shell Model, we discuss salient effects of the continuum coupling
on the one-neutron overlap integrals and the associated spectroscopic
factors in neutron-rich helium and oxygen nuclei. 
In particular, we demonstrate a characteristic near-threshold
energy dependence of the spectroscopic factors for different $\ell$-waves. 
We show also that the realistic radial overlap functions,
which are needed for the description of  transfer reactions, can be
generated by  single-particle wave functions of the appropriately chosen
complex  potential.
\end{abstract}


\end{frontmatter}

\section{Introduction}\label{intro}

Today, much interest in various fields of physics is devoted to the
study of small, open quantum systems (OQS), whose properties are profoundly
affected by the environment, i.e., continuum of decay channels. Although
every finite fermion system has its own characteristic features,
resonance phenomena are generic; they are great interdisciplinary
unifiers.  Many of these phenomena
have been originally studied in nuclear reactions, and now they are
explored in molecules in strong external fields, quantum dots and wires,
and other solid state micro-devices, crystals in laser fields, and
microwave cavities.

In the field of nuclear physics, a growing interest in the theory of OQSs
is associated with experimental efforts in producing
weakly bound and unbound nuclei close to the particle drip lines, and
studying structures and reactions with those exotic systems.  
In this context, the major challenge for nuclear theory is to
develop theories that would allow to understand properties of those
exotic  physical systems possessing  new and different properties
\cite{doba,brown,opr}. To this end, a unification of structure and
reaction aspects of weakly bound or unbound nuclei,  based on  the OQS
 formalism, is called for.

The nuclear shell model (SM) is the cornerstone of our understanding of
nuclei. In its standard realization \cite{brown,cau}, the SM assumes that
the many-nucleon system is perfectly isolated from an external
environment of scattering states and decay channels. The validity of
such a closed quantum system (CQS) framework is sometimes justified by
relatively high one-particle (neutron or proton) separation energies in
nuclei close to the valley of beta stability. However, weakly bound or
unbound nuclear states cannot be treated in a CQS formalism.  A
consistent description of the interplay between scattering states,
resonances, and bound states in the many-body wave function requires an
OQS formulation (see Ref.~\cite{opr} and references quoted
therein). Properties of unbound states lying above the particle (or
cluster) threshold directly impact the continuum structure. Coupling to
the particle continuum is also important for weakly bound states, such
as halos. A classic example of a threshold effect is the Thomas-Ehrman
shift \cite{te}, which manifests itself in the striking asymmetry in the
energy spectra between mirror nuclei. (See Ref.~\cite{ppnp} for more
discussion and examples related to this point.)

In this work, we investigate the impact of the non-resonant continuum on
single-nucleon overlap integrals and spectroscopic factors in weakly
bound and unbound nuclear states. Our paper is organized as  follows. 
The model used is the SM in the complex momentum-plane, the so-called
Gamow Shell Model (GSM) \cite{Mic02,Mic04,Bet02}, which is briefly
presented in Sec.~\ref{GSM_sec}. Section~\ref{GSM_H_spaces} describes
the concrete realization of the GSM employed in this work, i.e.,
interactions used and  the GSM model space, as well as the approximate
schemes that  have been introduced to illuminate specific physics points. 
The discussion of spectroscopic factors and overlap integrals is
contained in Sec.~\ref{spec_fact}. 
We show that the realistic radial
overlap functions, which are needed for the description of  transfer
reactions, can be generated by single-particle wave functions of the
appropriately chosen {\em complex average potential} reproducing the complex
$Q$-value of the studied reaction or decay. 
Furthermore, we demonstrate that the scattering continuum is of crucial
importance for the spectroscopic factors in the vicinity of
the particle emission threshold. This problem has been
discussed recently in our work~\cite{Mic07} for the $\ell=1$ neutron waves.
Here, we extend this discussion to $\ell=2$ and also
compare the  spectroscopic
factors in GSM with those calculated in the pole approximation to see
the influence of non-resonant continuum. As an illustrative
example  in these studies, 
we choose the case of {\em model}
two-neutron systems
outside the inert core:  $^{6}$He and $^{18}$O. Our aim is not to fit
the actual experimental data, but rather to illustrate generic effects.
Finally, the conclusions of this work are summarized in Sec.~\ref{conclusion}.

\section{Gamow Shell Model} \label{GSM_sec}

As the Gamow Shell Model has been described in detail in a number of
publications, only rudimentary information is provided here. In the 
roots of GSM lies the Berggren one-body completeness relation
\cite{Berggren1,Berggren2} that provides the mathematical foundation for
unifying   bound and unbound states. The Berggren ensemble consists of
bound, resonant, and scattering s.p. wave functions, generated by a
finite-depth potential $V(r)$, either real or complex. The wave
functions are regular solutions of the s.p. Schr{\"o}dinger equation,
\begin{equation}
u_\mathcal{B}''(r) =\left[ \frac{\ell(\ell+1)}{r^2} +
\frac{2m}{\hbar^2} V(r) - k^2 \right] u_\mathcal{B}(r),
\end{equation}
that obey outgoing or scattering boundary conditions:
\begin{equation}\label{one_body_state}
u_\mathcal{B}(r)_{r \rightarrow +\infty} =  C_+ \; H^+_\ell (kr) + C_- \; H^-_\ell (kr),
\end{equation}
where $k=\sqrt{2mE}/\hbar$ is the complex wave number.
 For the resonant states,  $C_-$=0 (outgoing boundary
condition).  In this paper, as we consider only valence neutrons,
$H^{\pm}_\ell$ is a Hankel  function. Normalization constants $C_+$
and $C_-$ are determined from the condition  that the radial wave
functions $u(r)$ are normalized to unity (for resonant states) or to the
 Dirac delta (for scattering states).

For a given partial wave $(\ell,j)$, the scattering states are
distributed along the contour $L_+^{\ell_j}$ in the complex momentum
plane. The set $|u_\mathcal{B}\rangle$  of all bound and resonant states  enclosed
between $L_+^{\ell_j}$ and the real $k$-axis, and scattering states
is complete  \cite{Berggren1}:
\begin{equation}
\int\hspace{-1.4em}\sum_{\mathcal{B}} |u_\mathcal{B} \rangle \langle
\widetilde{u_\mathcal{B}}| = 1.
\label{one_body_compl_rel}
\end{equation}
The Gamow (and Berggren) states are vectors in 
the rigged Hilbert space (or 
Gelfand triplet) \cite{Gel64,Bohm,Madrid,CivGad}. Consequently, in the above, 
the tilde symbol
 signifies that the complex
 conjugation arising in the dual space affects only the angular part and
 leaves the radial wave function  unchanged \cite{Berggren1,Berggren2}.
 The Berggren states are normalized using the squared radial
 wave function and not the modulus of the squared radial
 wave function. In the standard SM treatment, the latter normalization is used.

In numerical applications, the integral over scattering states along $L_+^{\ell_j}$
has to be discretized and the selected scattering states have to be 
renormalized \cite{Mic02}.
This leads to a discrete completeness relation:
\begin{eqnarray}
\sum_{\mathcal{B}=1}^{N} |u_{\mathcal{B}} \rangle \langle \widetilde{u_{\mathcal{B}}}| \simeq 1
~~;~~~~|u_{\mathcal{B}} \rangle = \sqrt{\omega_{\mathcal{B}}} \;
|u_{k_{\mathcal{B}}} \rangle,
\label{one_body_compl_rel_discr}
\end{eqnarray}
where $\{k_{\mathcal{B}},\omega_{\mathcal{B}}\}$ is the set of
discretized complex wave numbers and associated weights provided by a Gauss-Legendre
quadrature.  As discussed in Ref.~\cite{HagenVaagen}, the
 Gauss-Legendre integration formula  offers an excellent
 precision at a modest number ($\sim$30) of discretization points.

The many-body Berggren basis is that of Slater determinants  built
from s.p. states of Eqs.~(\ref{one_body_state}-\ref{one_body_compl_rel_discr}):
 \begin{eqnarray}
 &&|SD_i \rangle = | u_{i_1} \cdots u_{i_A} \rangle,  \label{SD}
 \end{eqnarray}
where the index $i$ labels the many-body basis, and $u_{i_j}$ is the
$j$-th s.p.~state occupied in $|SD_i \rangle$. The many-body
 completeness relation is a consequence of
Eq.~(\ref{SD}); it is obtained by forming all possible many-body Slater determinants:
 \begin{eqnarray}
 &&\sum_{i} |SD_i \rangle \langle \widetilde{SD_i}| \simeq 1. \label{Nbody_compl_rel_discr}
 \end{eqnarray}
The equality in Eq.~(\ref{Nbody_compl_rel_discr}) is not exact due to
the discretization of the one-body completeness relation.

In  the Berggren representation,  the SM  Hamiltonian matrix $H$ becomes
complex symmetric. Its complex-energy eigenvectors can be obtained by
using  the complex Lanczos method  \cite{diagnumbook}. The fundamental
difference between GSM and the real-energy SM is that the many-body
resonant states of the GSM  are embedded in the background of scattering
eigenstates, so that one needs a criterion to isolate them. The overlap
method introduced in Ref.~\cite{Mic02} has proven to be very efficient
in this respect. To this end, one  diagonalizes  $H$ in the basis
spanned on the s.p. resonant states only  (the so-called pole
approximation); this yields a zeroth-order approximation
$|\Psi_0\rangle$ for the wave function. The vector $|\Psi_0\rangle$ has a
correct outgoing behavior; it is used as a  pivot to generate a Lanczos
subspace of the full GSM space. The requested  resonant eigenstate of
$H$ is the one which maximizes the overlap $|\langle \Psi_0 |
\Psi\rangle|$. The overlap method has also been employed in the Density
Matrix Renormalization Group (DMRG) technique \cite{dmrg1} recently generalized
to treat the non-hermitian GSM case \cite{dmrg2}.

The definition of observables in GSM follows directly from the
mathematical setting of quantum mechanics in the rigged Hilbert space
\cite{Gel64,Bohm,Madrid,CivGad} rather than  the usual Hilbert space. A modified
definition of the dual space, embodied by the tilde symbol above the bra
states in Eqs.  (\ref{one_body_compl_rel}-\ref{Nbody_compl_rel_discr}),
implies that observables in many-body resonances become complex. In this
case, the real part of a matrix element corresponds to the expectation
value,  and the imaginary part can be interpreted as the uncertainty in
the determination of this expectation value due to the possibility of
decay of the state during the measuring process \cite{Berggren1,CivGad,Civ99}.
That is a standard price for replacing the time-dependent description of
an unstable system by the quasi-stationary description.

\section{Gamow Shell Model Implementation} \label{GSM_H_spaces}

This section contains details pertaining to GSM calculations
 carried out in this work. It describes the Hamiltonian used, the
 choice of the active Hilbert space, and the approximations employed.
 All remaining details of the formalism can be found in Refs. 
 \cite{Mic02,Mic07}.

 \subsection{GSM Hamiltonian}

 The doubly magic nuclei $^4$He and  $^{16}$O are assumed to be the
inert  cores in our  helium and oxygen calculations, respectively.
 The GSM Hamiltonian  consists of a one-body Woods-Saxon (WS)
 potential, representing a core-valence interaction,
 and  a two-body residual interaction:
 \begin{equation}
 H = \sum_{i} \left[ \frac{p_i^2}{2m} + V_{WS}(r_i) \right]  + V^{(2)}_{res}.
 \end{equation}
 The WS potential contains the central term and the spin-orbit term:
 \begin{equation}
 V(r) = - V_{WS} \cdot f(r) - 4 \; V_{so} \left( \vec{l} \cdot
 \vec{s} \right) \frac{1}{r} \left| \frac{df(r)}{dr} \right|,
 \label{WS_pot}
 \end{equation}
 where  $V_{WS}$ and $V_{so}$ are  the  strength constants and
 \begin{equation}
 f(r) = \left[ 1 + \exp \left( \frac{r-R_0}{d} \right) \right]^{-1}
 \end{equation}
 is the  WS form-factor characterized by radius $R_0$ and
 diffuseness $d$. In  the helium calculations, we employed
 the ``$^{5}$He" WS parameter set \cite{Mic02} which reproduces the
 experimental energies and widths of known  s.p. resonances $3/2_1^-$ and
 $1/2_1^-$ in $^5$He ($0p_{3/2}$ and $0p_{1/2}$
 resonant states  in our model). In
 the oxygen case, we used
 the ``$^{17}$O" WS parameter set \cite{Mic02}  that
 reproduces experimental energies and widths of the
 $5/2_1^+$ and $1/2_1^+$ bound states and the $3/2_1^+$ resonance in
 $^{17}$O (i.e., $0d_{5/2}$, $1s_{1/2}$, and $0d_{3/2}$  resonant states).

 We used a different residual interaction for helium and oxygen isotopes.
 This choice has been motivated by our previous studies of oxygen, helium,
 and lithium chains \cite{Mic02,Mic04}.
 For the heliums,  we took
 the  finite-range surface Gaussian interaction (SGI) \cite{Mic04}
 \begin{equation}
 V^{(2)}_{res} = \sum_{i < j} V_0^{(J)}
 \cdot \exp \left(-\left[ \frac{\vec{r}_i - \vec{r}_j}{\mu}
 \right]^2 \right) \cdot \delta(|\vec{r}_i| + |\vec{r}_j| - 2 R_0)
      \label{H_He}
 \end{equation}
 with  the range $\mu$=1\,fm and the coupling constants depending on
 the total angular momentum $J$ of the neutron pair: $V_0^{(0)}=-403$ MeV
 fm$^{3}$, $V_0^{(2)}=-392$ MeV fm$^{3}$. These constants are fitted to
 reproduce the ground-state (g.s.) binding energies of $^6$He and $^7$He in GSM.

 For the oxygens, we employed a surface delta interaction (SDI)
 \cite{Mic04}:
 \begin{equation}
 V^{(2)}_{res} = \sum_{i < j} V_0 \cdot \delta \left( \vec{r}_i - \vec{r}_j
 \right) \cdot \delta(|\vec{r}_i| - R_0), \label{H_O16_O17}
 \end{equation}
 where the SDI coupling constant $V_0$=--700 MeV fm$^3$
 was fitted to
 in order to reproduce the two-neutron separation energy of
 $^{18}$O.

 \subsection {GSM Configuration Space}

In the helium variant,  the  valence space
consists of the $p_{3/2}$ and $p_{1/2}$ neutron partial waves. The
$p_{3/2}$ wave functions include a $0p_{3/2}$ resonant state and
$p_{3/2}$ non-resonant scattering states along a complex contour
enclosing the $0p_{3/2}$ resonance in the complex $k$-plane. For a
$p_{1/2}$ part, we take non-resonant scattering states along the
real-$k$ axis (the broad $0p_{1/2}$ resonant state  plays a negligible
role in the g.s.~wave function of $^{6}$He). For both contours, the
maximal momentum value is $k_{\rm{max}}=$3.27 fm$^{-1}$. The contours
have been discretized with up to 60 points,  and the attained precision on
energies and widths is better than 0.1 keV. 

In the oxygen case, the valence GSM space for neutrons consists of the
$0d_{5/2}$, $1s_{1/2}$, and $0d_{3/2}$ Gamow states  and the
non-resonant complex $d_{3/2}$ continuum. The $0d_{5/2}$ and  $1s_{1/2}$
shells are bound while the $0d_{3/2}$ state is a narrow resonance. The
maximum value for $k$ on the contour $L_+^{d_{3/2}}$ is
$k_{\rm{max}}=$1.5 fm$^{-1}$. The contour has been discretized with 45
points and the resulting precision on energies and widths is $\sim0.1$
keV. The real-energy $d_{5/2}$ and $s_{1/2}$ contours  give rise to a
renormalization of the residual interaction coupling constant; hence,
they have been neglected \cite{Mic02}.

 \subsection {Simplified SM schemes: GSM-p and HO-SM}

In order to illuminate certain aspects of GSM results (e.g.,
non-resonant continuum coupling or  configuration mixing), we have
introduced  two simplified  SM schemes: the GSM in the  pole
approximation (GSM-p) and the harmonic oscillator shell model (HO-SM).

In GSM-p, the scattering components of the Berggren basis are
disregarded, and only the resonant states  are present in the basis. In
particular, they will be the $0p_{3/2}$ and $0p_{1/2}$ neutron states
for heliums, and $0d_{5/2}$, $1s_{1/2}$, and $0d_{3/2}$ neutron states for
oxygens. In this case, the one-body completeness relation
(\ref{one_body_compl_rel}) is obviously violated. Still, the comparison
between full GSM results and GSM-p is instructive as it illustrates the
influence of the non-resonant continuum subspace on the GSM results.

For the HO-SM calculation,  the radial wave functions of the GSM-p basis
are replaced by those of the spherical harmonic oscillator (HO)  with
the frequency $\hbar \omega=41A^{-1/3}$ MeV. The real energies of the
resonant states define the one-body part of the HO-SM Hamiltonian. The
HO-SM scheme is intended to illustrate the ``standard"  CQS SM
calculations in which only bound valence  shells are considered in the
s.p. basis.

 \section{One-nucleon overlap integrals and spectroscopic factors in GSM} \label{spec_fact}

\subsection{Definitions}

Single-nucleon overlap integrals and the associated spectroscopic
factors  are basic ingredients of the theory of direct reactions
(single-nucleon transfer, nucleon knockout, elastic break-up)
\cite{satch,sfs,Glenden,FroLipp}.  Experimentally, spectroscopic factors  can be  deduced
from measured cross sections; they are useful measures of  the
configuration mixing in the many-body wave function. The associated
reaction-theoretical analysis often reveals model- and probe-dependence
\cite{sf1,sf2,sf3} raising concerns about the accuracy of  experimental
determination of spectroscopic factors. In our study we discuss the  uncertainty in
determining spectroscopic factors due to  the two assumptions commonly used in  the
standard SM studies, namely (i) that a nucleon is transferred to/from a
specific s.p.~orbit (corresponding to an observed s.p.~state), and  (ii)
that the transfer to/from the continuum of non-resonant scattering
states can be disregarded. This discussion complements our recent
work \cite{Mic07} whose focus was on threshold anomalies and channel coupling.

The one-nucleon radial overlap integral is discussed in the usual way
 \cite{sfs,Glenden,FroLipp}:
 \begin{equation}
 u_{\ell j}(r) = \langle \Psi^{J_{A}}_{A} (\vec{r}_1,\cdots,\vec{r}_A = \vec{r})
     | \left[ | \Psi^{J_{A-1}}_{A-1} (\vec{r}_1,\cdots,\vec{r}_{A-1})
      \rangle \otimes |\ell,j \rangle \right]^{J_A} \rangle, \label{ovf_basic_def}
 \end{equation}
where $|\Psi^{J_{A}}_{A}\rangle$ and $|\Psi^{J_{A-1}}_{A-1}\rangle$ are
wave functions of nuclei $A$ and $A-1$, respectively, and  angular and
spin coordinates (represented by $|\ell,j\rangle$)  are integrated out
so that $u_{\ell j}$ depends only on $r = |\vec{r}|$. The spectroscopic factor  is given
by the real part of the norm  $S^2$ of the overlap integral.
 Using a decomposition of the s.p.
$(\ell,j)$ channel in the complete Berggren basis, one obtains GSM
expressions for the overlap integral and $S^2$:
 \begin{eqnarray}
 &&u_{\ell j}(r) = \int\hspace{-1.4em}\sum_{\mathcal{B}} \langle \widetilde{\Psi^{J_{A}}_{A}} || a^+_{\ell j}(\mathcal{B}) || \Psi^{J_{A-1}}_{A-1} \rangle
 \mbox{ } \langle r \ell j | u_\mathcal{B} \rangle \label{ovf_GSM} \\
 &&S^2 \equiv \int u_{\ell j}^2(r)dr = \int\hspace{-1.4em}\sum_{\mathcal{B}} \langle \widetilde{\Psi^{J_{A}}_{A}} || a^+_{\ell j}(\mathcal{B}) || \Psi^{J_{A-1}}_{A-1} \rangle^2 \label{eq3}
 \end{eqnarray}
 where $a^+_{\ell j} (\mathcal{B})$ is a
 creation operator associated with
 a s.p.~basis state $|u_\mathcal{B}\rangle$.
We wish to emphasize that
since Eqs.~(\ref{ovf_GSM},\ref{eq3}) involve summation over all discrete
Gamow states and integration over all scattering states along  the
contour $L_+^{\ell_j}$, the final result is independent of the
s.p.~basis assumed (see also Ref.~\cite{model_indep_exp} where a
model-independent spectroscopic factor experimental extraction procedure is proposed).
This is in contrast to the standard  SM treatment of spectroscopic factors  where
model-dependence enters through the specific choice of a s.p.~state $|n
\ell j\rangle$ \cite{modeldep}, with Eq.~(\ref{eq3}) reducing to a
single  matrix element $\displaystyle|\langle \Psi^{J_{A}}_{A} || a^+_{n
\ell j} || \Psi^{J_{A-1}}_{A-1} \rangle|^2$.

Spectroscopic factors are often extracted from measured transfer reaction cross sections.
(In the context of the following discussion,  is worth noting that while the
reaction cross sections are measurable quantities, the spectroscopic factors are purely
theoretical concepts. First, they are deduced from experiment in a model-dependent way.
Second,  occupation numbers deduced from spectroscopic factors are not observables per se
\cite{Fur02}.)
In the CQS SM framework, the transfer total cross section is given by
\cite{Bohr,Glenden,FroLipp}:
 \begin{eqnarray}
 &&\sigma = \sum_{n \ell jm} S_{n \ell jm} \; \sigma_{s.p.}^{n \ell jm},
 \label{d_sigma}
 \end{eqnarray}
where $S_{n \ell jm}$ is the spectroscopic factor associated with the $|n \ell jm \rangle$
s.p. state, and $\sigma_{s.p.}^{n \ell jm}$ a s.p. cross section
containing all the kinetic dependence related to the transfer of the
nucleon of the projectile to the shell model state. While 
Eq.~(\ref{d_sigma}) is, in principle, exact for bound s.p. states, as one
sums over all quantum numbers $n \ell jm$, the commonly used
approximation is to take only one radial quantum number $n$ for each
$\ell jm$ in the sum of Eq.~(\ref{d_sigma}), i.e., one assumes that the
nucleon is transferred to/from an orbit with a specific radial quantum
number.  Hence, the resulting value of spectroscopic factor becomes spuriously
basis-dependent. We shall show in the following that this can lead to
sizeable errors if  the  states $|\Psi^{J_{A}}_{A}\rangle$
or $|\Psi^{J_{A-1}}_{A-1}\rangle$ are
loosely bound. 

Obviously, the factorization (\ref{d_sigma}) does not apply to many-body
states which cannot be described in the CQS formalism. Firstly, in the
OQS formalism the sum in Eq.~(\ref{d_sigma}) is not limited to
discrete states $|n \ell j m\rangle$ only. The appearance of the 
cusp at the channel threshold both in the reaction cross-section and in
the real part of the GSM spectroscopic factor suggests an expression:
 \begin{eqnarray}
 &&\sigma = \sum_{\ell j} Re(S_{n \ell j}) \; \sigma_{s.p.}^{\ell j},
 \label{d_sigma1}
 \end{eqnarray}
as the relation between the observable reaction cross-section and the
real part of the spectroscopic amplitude $S\equiv S_{n \ell j}$ (cf
(\ref{eq3})) calculated in GSM. Secondly, as stated earlier,
observables involving unbound states are complex in GSM. It was
conjectured by Berggren \cite{Berggren1} and later by Gyarmati {\it et al.}
\cite{Gyarmati} that the real part   of a matrix element corresponds to
the expectation value. Based on  scattering theory, Berggren
showed \cite{Berggren3} that the real part of the complex cross section
for populating a resonance is equal to the energy integral of the
in-elastic continuum cross section across the resonance peak. The
imaginary part of the cross section can be  identified with the strength of
the resonance-background interference. This conclusion was later
generalized to hold for any observable operator involving resonant
states. In this work, we follow Berggren's interpretation and associate
an imaginary part of a matrix element (e.g., spectroscopic factors or
radial overlap function) with the uncertainty in the determination of
the real part of this matrix element.

For weakly bound or unbound nuclei, the transfer
reaction cross section is changed by couplings to the non-resonant
continuum which modify both the effective interaction among valence
particles and the configuration mixing. This change in the cross section
has a regular component having a smooth dependence on the distance from
the particle-emission threshold, and an anomalous component which
results in  the Wigner-cusp phenomenon \cite{Wigner}. Other quantities,
such as  the strength function \cite{Hategan1,Graw}, the
continuum-coupling correction to the CQS eigenvalue \cite{Op05}, or the
overlap integral \cite{Mic07} are modified in a similar way. 
These two components can be seen by comparing
full GSM results with equivalent SM results (here: the HO-SM approximation).

 Generic features of wave functions in the vicinity of the reaction
 threshold (the particle-emission threshold) are responsible for the
 anomalous component. Wigner was first to notice it  as the
 characteristic behavior of scattering and reaction cross-sections
 \cite{Wigner}, often referred to as the Wigner threshold law or the
 Wigner-cusp phenomenon. A quantitative explanation of this behavior of
 cross sections was given later
 \cite{Breit,Baz,Newton,Meyerhof,Baz1,Lane}. Wigner threshold law tells
 that  in the vicinity of particle-emission thresholds the reaction
 cross section has a universal shape depending on the charge and the 
 orbital angular momentum of the scattered particle. In particular, the
 reaction cross section for neutron scattering near the neutron emission
 threshold shows a characteristic cusp dependence on the orbital
 momentum of a scattered neutron. This salient $\ell$-dependent
 anomalous behavior (cusp) is absent for charged scattered particles.

Contrary to the universality of the cusp, its magnitude is not a generic
feature of OQSs and cannot be found from general principles. Also the
ratio between regular and anomalous components in the near-threshold
behavior of reaction cross-sections, strength functions, radial overlap
integrals, spectroscopic factors, or continuum-coupling correction to CQS eigenvalues is
the specific property of each considered OQS and its Hamiltonian.

The modification of the cross section in one channel manifests itself
due to the unitarity of the scattering matrix in other open channels.
This coupling effect has been studied theoretically
\cite{Baz,Newton,Meyerhof,Baz1,Hategan,Hategan1,Mic07} and
experimentally \cite{Malmberg,Wells,Moore,Abramovich} in connection with
the Wigner threshold anomaly. The channel-channel continuum coupling 
influences both the anomalous and regular cross-section components.
Similar multi-channel threshold effects are  expected  to show up in
various observables which depend on occupation probabilities of s.p.
shells and the configuration mixing involving continuum states.

It was shown in Ref.~\cite{Mic07} that once a given nuclear Hamiltonian
is chosen, quantities such as the spectroscopic factors, defined in GSM through the norm
of the overlap integral, and their behavior close to particle-emission
thresholds are defined uniquely in terms  of the exact many-body GSM
solutions. Below, we shall demonstrate that the threshold variations due
to the continuum coupling can be of a comparable size to those generated
by the long-range correlations and the coupling to high-momentum states
reached by short-range and tensor components of the nucleon-nucleon
interaction \cite{Dic04}. Hence, the states of {\em open} and {\em
closed} quantum systems may have not only different asymptotic behavior
 but also  different shell model structure. In that sense,
loosely bound or unbound states form a different class of correlated
quantum many-body systems than those found in well-bound nuclear states.

\subsection{The $^{5}$He+n$\rightarrow$$^{6}$He case}  \label{He_sec}

We first consider the g.s. spectroscopic factor of \mbox{ }$^6$He corresponding to the
channel $[{^5}{\rm He}({\rm g.s.})\otimes p_{3/2}]^{0^+}$. In order to
investigate the effect of the continuum coupling, the depth of the WS
potential is varied so that the energy $e_{p_{3/2}}$ of the $0p_{3/2}$
s.p.~state (the lowest $p_{3/2}$ pole of the $S$-matrix) changes its
character from bound to unbound. In our model space, the energy
$e_{p_{3/2}}$ is both  the g.s. energy of $^5$He and (negative of) the
one-neutron (1n) separation energy $S_{1n}$ of  $^5$He.
 \begin{figure}
 \centering
 \includegraphics*[width=12cm]{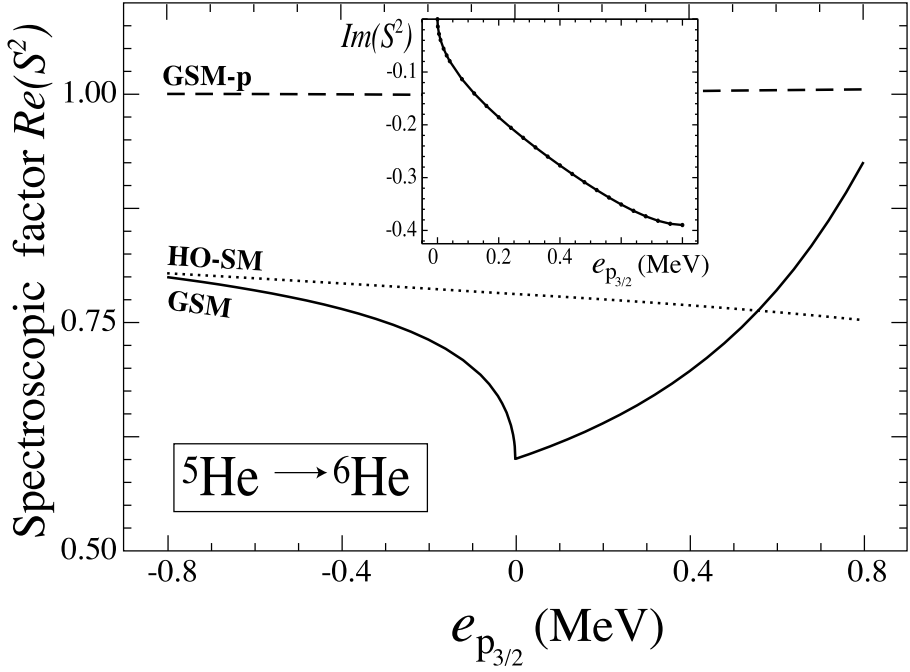}
\caption{The real part of the overlap $\displaystyle \langle 
^6\rm{He(g.s.)} | [^5{\rm He(g.s.)} \otimes p_{3/2}]^{0^+} \rangle^2$ 
 as a function of the energy of the $0p_{3/2}$
resonant state.  $^6$He is always bound. 
  The solid line (GSM) shows  the full GSM result. The
dotted line (HO-SM) corresponds to the SM calculation in the oscillator
basis of $0p_{3/2},0p_{1/2}$. The dashed line (GSM-p) shows the  GSM
result in the pole approximation. The imaginary part of $S^2$  is given
in the inset. In order to better illustrate the effect of configuration
mixing, $S^2$ has been normalized  to 1 in the limit of vanishing  
residual interaction. }
 \label{fig_01}
 \end{figure}

The calculated spectroscopic factor [$^5$He(g.s.)$\otimes$$p_{3/2}]^{0^+}$ in $^6$He(g.s.)
is shown in Fig.~\ref{fig_01} as a function of the real energy $e_{p_{3/2}}$
of the $0p_{3/2}$ pole.  In the range of $e_{p_{3/2}}$ values shown in
Fig. \ref{fig_01}, $^6$He is bound ($S_{1n}$($^6$He)$>$0). The spectroscopic factor strongly depends
 on the position of the pole: for $e_{p_{3/2}}$$<$0 
 ($S_{1n}$($^5$He)$>$0)
  it
decreases with $e_{p_{3/2}}$ while  it  increases  for $e_{p_{3/2}}$$>$0
 ($S_{1n}$($^5$He)$<$0). At the 1n-emission threshold in $^5$He ($e_{p_{3/2}}$=0), 
spectroscopic factor exhibits a cusp. At this point, the derivative of spectroscopic factor becomes 
discontinuous and the coupling matrix element between the resonant
$0p_{3/2}$ state  and the non-resonant continuum reaches its maximum
\cite{Op05}.

As discussed in Ref.~\cite{Mic07}, the dependence of spectroscopic factor on $e_{p_{3/2}}$
around $e_{0p_{3/2}}$=0 (where the spectroscopic factor exhibits a cusp due to a 
coupling with the $^4$He+n+n channel)
follows the  threshold behavior of the reaction cross
section  \cite{Wigner}. Specifically, the anomalous component of spectroscopic factor {\it
below} the 1n threshold  in $^5$He behaves as $(-e_{\ell
j})^{\ell-1/2}$.  {\it Above} the  threshold, the spectroscopic factor factor is complex;
the real part behaves as $(e_{\ell j})^{\ell+1/2}$ while the imaginary
part, associated with the decaying nature of $^5$He, varies as
$(e_{\ell j})^{\ell-1/2}$. 

To assess the role of the  continuum, both resonant and non-resonant,
the GSM results are compared in  Fig.~\ref{fig_01} with the HO-SM and
GSM-p calculations. In contrast to GSM,   spectroscopic factors in HO-SM and GSM-p
 vary  little in the 
energy range considered and no threshold effect is seen. It is interesting to note
that in the limit of an appreciable binding, the $0p_{3/2}$ wave
function is fairly well localized, the importance of the continuum
coupling is diminished, and the HO-SM result approaches the GSM limit.
For $e_{p_{3/2}}>0$, GSM results are in strong disagreement
with both HO-SM and GSM-p. A $\sim$25\% difference between spectroscopic factors in HO-SM
and GSM-p  is  striking. This difference reflects a
strong dependence of the two-body matrix elements 
 on the actual radial wave functions. In the GSM-p variant,
the g.s. of $^6$He is described by an almost  pure $[0p_{3/2} \otimes
0p_{3/2}]^{0^+}$ configuration and the resulting  spectroscopic factor is close to one in
the whole energy region considered. A  difference between GSM
and GSM-p results illustrates the impact of the non-resonant continuum.

 \begin{figure}[htb]
 \centering
 \includegraphics[width=9cm]{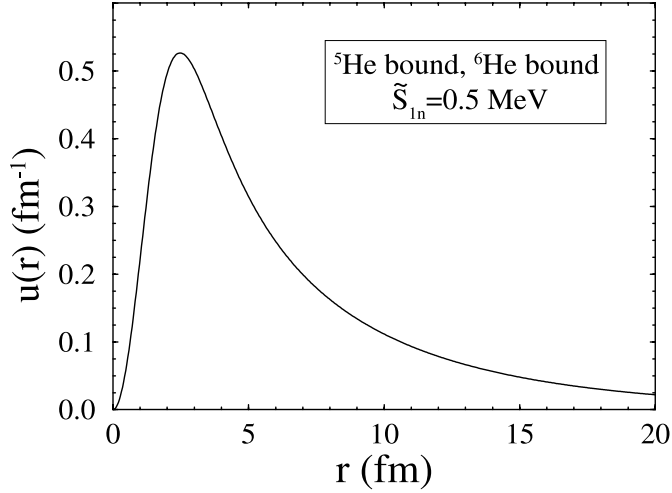}
 \caption{Single-neutron radial overlap integral $u(r)$ for
the   $\displaystyle \langle ^6{\rm He( g.s.}) | [^5{\rm He( g.s.})
 \otimes p_{3/2}]^{0^+} \rangle^2$  channel 
 calculated in GSM for one-neutron bound
 $^5$He and  $^6$He. The radial wave function of the
 $0p_{3/2}$ resonant state of the {\em real} WS potential adjusted
 to reproduce the GSM value of $S_{1n}$ in  $^6$He practically coincides 
with $u(r)$. The overlap integral has been 
 normalized to unity to allow comparison with the WS wave function.}
 \label{case1}
 \end{figure}

\begin{figure}[htb]
 \centering
 \includegraphics[width=9cm]{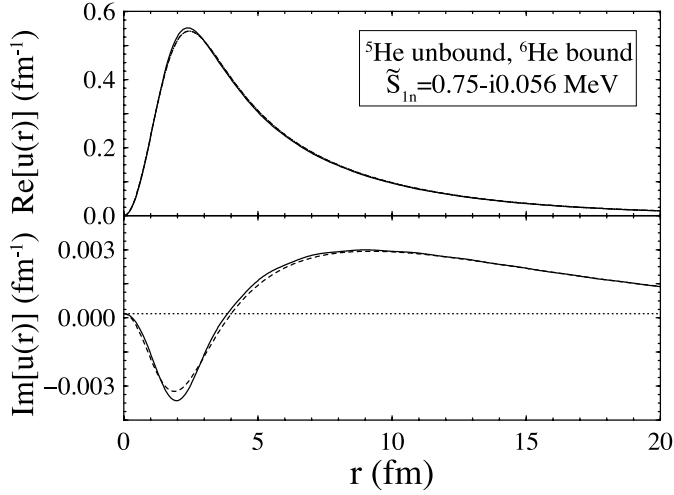}
 \caption{Solid line:  the single-neutron radial overlap integral $u(r)$
 for the   $\displaystyle \langle ^6{\rm He( g.s.}) | [^5{\rm He( g.s.})
 \otimes p_{3/2}]^{0^+} \rangle^2$  channel calculated in GSM for
 unbound $^5$He  and bound $^6$He. Dotted line: the radial wave function
 of the $0p_{3/2}$ resonant state of the {\em real} WS potential whose
 strength was adjusted to reproduce the GSM value of $S_{1n}$ in $^6$He.
 Dashed line: the radial wave function of the $0p_{3/2}$ resonant state
 of the {\em complex} WS potential whose complex strength was  adjusted
 to reproduce the GSM value of the generalized separation energy
 ${\tilde S}_{1n}$ in $^6$He. All wave functions are normalized to unity
 to allow comparison. The real (imaginary) parts of the wave function
are  shown in the top (bottom) panel.}
 \label{case4}
 \end{figure}

\begin{figure}[htb]
 \centering
 \includegraphics[width=9cm]{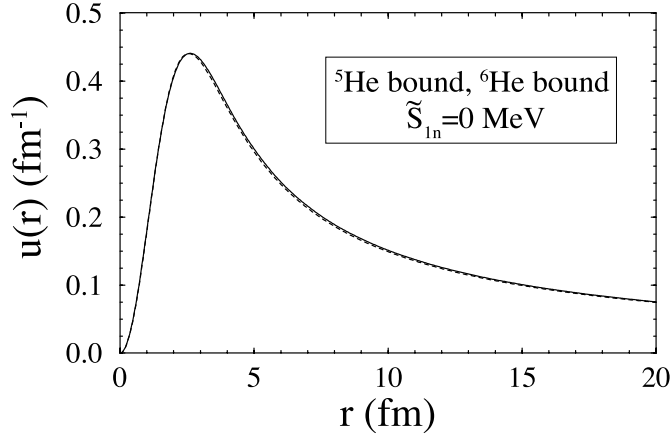}
 \caption{Similar to Fig. \ref{case1}  except for the threshold situation of
 $S_{1n}$[$^6$He]=0.}
 \label{case7}
 \end{figure}

 Figures~\ref{case1}-\ref{case7} display one-neutron radial overlap integrals
 calculated in GSM for bound $^6$He 
 and three energies of the $0p_{3/2}$
 pole corresponding to different physical situations considered in
 Fig. \ref{fig_01}. 
 In Fig. \ref{case1}, both $^5{\rm He}$ and $^6{\rm He}$
 are bound with one-neutron separation
 energies 0.5 MeV and 1 MeV,
 respectively. 
Figure~\ref{case4} illustrates the case of unbound $^5{\rm He}$ 
 ($E[^5$He]=0.25 --i0.056 MeV) and bound $^6{\rm He}$ ($E[^6$He]=--0.5 MeV).
Here, the generalized one-neutron separation energy, $ {\tilde S}_{1n}$,
defined as a difference
 between g.s. binding energies of neighboring nuclei, becomes {\em complex}:
 \begin{equation}\label{genSn}
 {\tilde S}_{1n}(N)\equiv E(N-1)-E(N) =  S_{1n}(N) -{i\over 2} \left[
 \Gamma(N-1)-\Gamma(N)
 \right].
 \end{equation}
The real part of ${\tilde S}_{1n}$ is the usual separation energy $S_{1n}$
while the  imaginary part appears when
either parent or daughter nucleus is unbound.
 %
 Finally, Fig. \ref{case7} shows $u(r)$
for the case of bound $^5$He and $^6$He 
 having the same energy, i.e., $S_{1n}$[$^6$He]=0.
 %

 The overlap integral is peaked around the surface ($\sim$2\,fm)  and
 its asymptotic behavior is determined by  one-neutron separation energy 
 \cite{Blo77,Tim03}:
 \begin{equation}\label{uas}
 u(r) \rightarrow e^{-\kappa r},
 \end{equation}
 where the decay constant is
 \begin{equation}\label{uas1}
\kappa = \sqrt{2m S_{1n}}/\hbar.
 \end{equation}
In the case considered, $u(r)$ is expected to decay exponentially with
$\kappa$  determined by the one-neutron  separation energy of $^6$He.
Figures~\ref{case1}-\ref{case7} show, by the dotted line, the radial wave
function of the $0p_{3/2}$ resonant state of the WS potential with a
depth adjusted to reproduce the GSM value of $S_{1n}$ in $^6$He. Since
$^6$He is one-neutron bound, the  corresponding $0p_{3/2}$ WS state
is bound; hence,  its  wave function is real. It is seen that the agreement
between  the WS wave function and the real part of $u(r)$  is
excellent. 
 
The situation shown in  Fig.~\ref{case4} is particularly
interesting  as
the overlap integral is complex due to an unbound nature of
$^5$He. Here, a real WS potential cannot provide any information
about the imaginary part of $u(r)$. One possible
way of overcoming this difficulty is to introduce
a {\em complex} WS potential characterized by a complex
strength:
\begin{equation}
V_{WS} \equiv V^R_{WS} +i V^I_{WS}.
\end{equation}
Following the usual SM strategy applied to real $u(r)$ cases, $V_{WS}$ can be
adjusted to reproduce the generalized separation energy (\ref{genSn}).
This guarantees that the asymptotic behavior of the overlap integral,
given by Eqs.~(\ref{uas}-\ref{uas1}) with $S_{1n}$$\rightarrow$$\tilde{S}_{1n}$,
 is correct. (The extension of the
discussion of asymptotic properties of $u(r)$ in Ref.~\cite{Blo77} to
the complex-energy case is straightforward; it follows  from the
analytic properties of the vertex form factor.)
It is seen  in Fig.~\ref{case4} (dashed line) that
both real and imaginary parts of the  radial overlap
integral are well reproduced by the $0p_{3/2}$  resonant  wave
function of a WS potential that reproduces the GSM value of
$\tilde{S}_{1n}$. This  $0p_{3/2}$ complex-energy state
corresponds to a  `decaying bound state' which can be found only in the
complex s.p. potential (see, e.g., \cite{Bay96,Cha01}); asymptotically,
such a state exhibits an exponentially damped oscillation.
The effect of the non-resonant continuum is seen in a
slightly better localization of the GSM overlap. 

Figure~\ref{case7} illustrates the threshold limit. 
Here, $S_{1n}$=0 and $u(r)$ shows asymptotic behavior that is not
exponential. Also in this case
the GSM radial overlap integral is well reproduced by the
radial $0p_{3/2}$ WS wave function corresponding to a halo
state located at  zero energy.

The Wigner estimate for the near-threshold behavior of cross section
\cite{Wigner} is independent of the reaction mechanism. While the Wigner
limit has been  reached in Fig.~\ref{fig_01} by changing  the pole of
the one-body S-matrix, it is interesting to see whether a similar
pattern  can be generated by many-body correlations.
 \begin{figure}
 \centering
 \includegraphics[width=10cm]{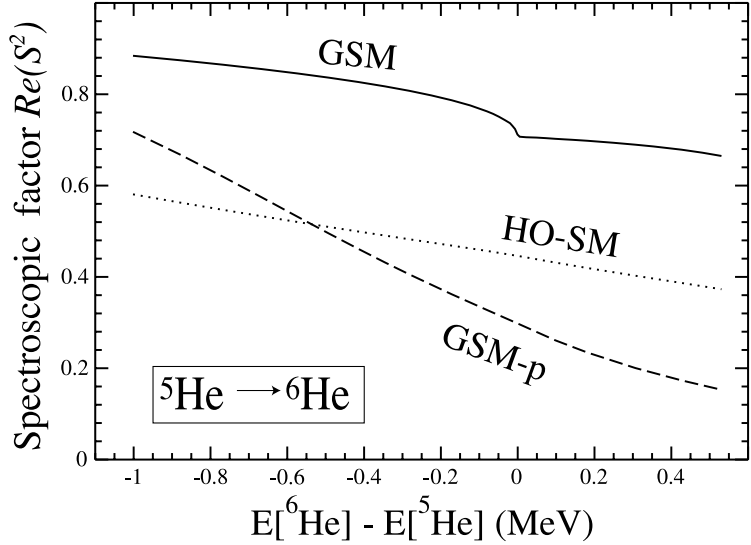}
 \caption{Similar to Fig.~\ref{fig_01} except as a function of
 (negative)  $S_{1n}$
 of $^{6}$He. The depth of the WS potential of $^{5}$He has been
 adjusted  to bind  $^{5}$He (see text). The separation energy $S_{1n}$
 of $^6$He has been varied by changing  the  coupling constant
 $V_0^{(J=0)}$ of the SGI two-body interaction.} \label{fig_03}
 \end{figure}
Figure~\ref{fig_03} illustrates a direct case of the Wigner cusp. Here,
we have fixed the  WS potential so that the $0p_{3/2}$
 and $0p_{1/2}$ shells are both bound with respective energies  of $-5$\,MeV and
 $-0.255$\,MeV, and we have varied the SGI coupling constant
 $V_0^{(J=0)}$ so that  $S_{1n}$[$^6$He] changes sign. 
 The behavior of spectroscopic factor around the 1n threshold  is similar to that  of
 Fig.~\ref{fig_01}, with the GSM 
spectroscopic factor  exhibiting a  non-analytic behavior at $S_{1n}$=0. This result
 constitutes an excellent test of the GSM formalism: the Wigner limit
 is reached precisely at a threshold obtained from many-body
 calculations. In the whole range of considered separation energies
 the HO-SM and GSM-p results vary  smoothly
and they differ from the GSM results.
 In this example, the anomalous component of spectroscopic factor
 is much smaller than the regular component showing the essential role
 played by the non-resonant continuum component  in the ground state 
 of $^6$He.

 \begin{figure}[htb]
 \centering
 \includegraphics[width=9cm]{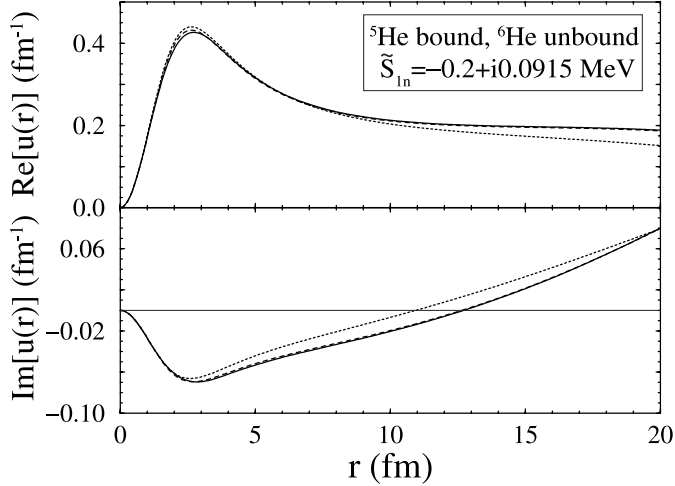}
 \caption{Similar to Fig.~\ref{case4}  except  for bound $^5$He and
 unbound  $^6$He ($S_{1n}$ of $^6$He is negative.) 
 The dashed line depicts the radial wave function of
 the $0p_{3/2}$ resonant state of the complex WS potential
 whose depth is adjusted to reproduce the GSM value of the generalized 
 separation energy ${\tilde S}_{1n}$ in $^6$He.} \label{case2}
 \end{figure}

\begin{figure}
 \centering
 \includegraphics[width=10cm]{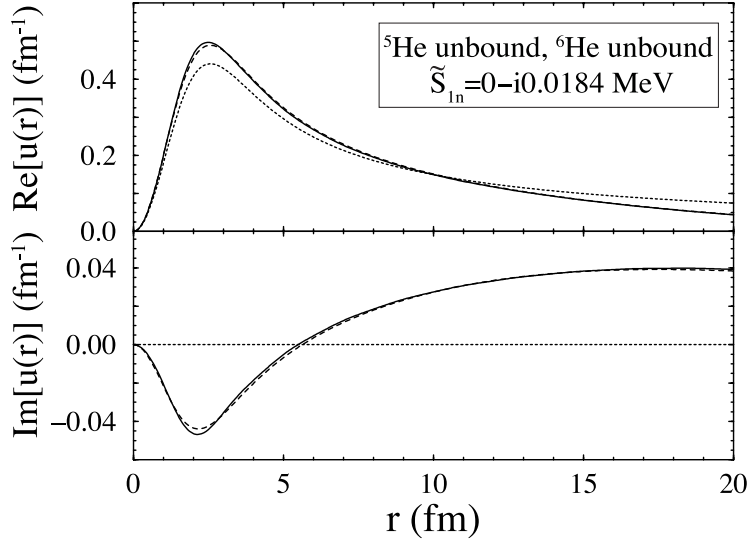}
 \caption{Similar to Figs. \ref{case4}  and \ref{case2} except
  for unbound
 $^5$He and  $^6$He. The 1n  separation energy $S_{1n}$ of $^6$He
has been adjusted to zero. 
The generalized 1n separation energy ${\tilde S}_{1n}$ has
a non-vanishing imaginary part. } \label{case8}
 \end{figure}

 Figures~\ref{case2} and \ref{case8} display $u(r)$
 calculated for the two cases shown in Fig.~\ref{fig_03}
 representing  unbound $^6$He, and  $^5$He being either bound 
 (Fig. \ref{case2}) or unbound (Fig. \ref{case8}). In
 Fig. \ref{case2}, the generalized separation energy 
 is complex, ${\tilde S}_{1n}$=--0.2+i0.0915\,MeV. The corresponding 
 Gamow pole is a bound decaying state.
 Figure~\ref{case8} shows the radial overlap integral for unbound
 $^5$He  and $^6$He having the same real 
 energies, i.e.,  $S_{1n}$=0, but different
widths.  Consequently, the generalized
 separation energy ${\tilde S}_{1n}$ has a  non-zero imaginary part in
 this case. The corresponding resonant state lies on the diagonal of the 
 fourth quarter of the complex-$k$ plane.

The radial wave functions of the
 $0p_{3/2}$ resonant state  of the real WS potential 
 adjusted to the GSM separation energy  of $^6$He are shown by
 the dotted lines in Figs. \ref{case2} and
 \ref{case8}. The agreement between  the 
  WS wave function and  $u(r)$ is 
 poor. In particular, their  real parts 
 have a  different asymptotic behavior.  In Fig. \ref{case8},
the imaginary part of the $0p_{3/2}$ WS wave function
 is  zero by construction, and it
 does not reproduce the salient behavior of the imaginary part of $u(r)$.
On the other hand, the 1n radial overlap integral is very well
reproduced by the $0p_{3/2}$ resonant state wave function
 of the complex WS potential (dashed curves in Figs. \ref{case2}, \ref{case8}).

In general, we find that shapes of  1n  overlap integrals calculated
 in GSM for any complex value of $\tilde{S}_{1n}$ are
 well reproduced by radial wave functions of a
complex WS potential whose depth is adjusted to reproduce the
 complex separation energy. In particular,
 for unbound parent and/or daughter nuclei, the
 asymptotic behavior of $u(r)$  is given by the generalized
 complex separation energy ${\tilde S}_{1n}$ and not by the 
 real separation energy $S_{1n}$.

 \subsection{The $^{17}$O+n $\rightarrow$  $^{18}$O case} \label{O_sec}
 \begin{figure}[htb]
 \centering
 \includegraphics*[width=12cm]{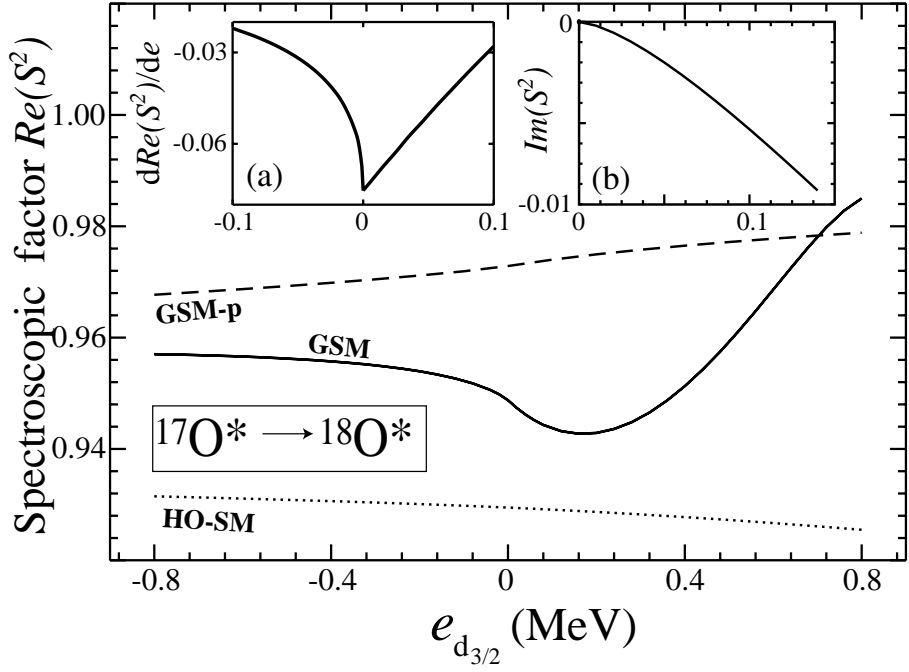}
 \caption{Similar to Fig.~\ref{fig_01} except for the overlap
 in the excited  $0^+_3$ state of $^{18}$O: $\displaystyle \langle
 ^{18}{\rm O}( 0^+_3) | [^{17}{\rm O}( 3/2_1^+) \otimes d_{3/2}]^{0^+}
 \rangle^2$. The first derivative  of the spectroscopic factor in the neighborhood of the
 $e_{d_{3/2}}$=0 threshold  is shown in the inset (a) while  inset (b) displays
 the imaginary part of $S^2$.}
 \label{fig_04}
 \end{figure}
 To investigate the dependence of spectroscopic factors on the orbital angular momentum
 $\ell$ of the transferred nucleon, we consider the $d_{3/2}$ partial
 wave  in $^{18}$O. We show in Fig.~\ref{fig_04} the spectroscopic factor for
 the excited  $0^+_3$ state of $^{18}$O in the channel
 $[^{17}{\rm O}(3/2_1^+) \otimes d_{3/2}]^{0^+}$.
 The behavior of spectroscopic factor shown
 in Fig.~\ref{fig_04} is similar to that of Fig.~\ref{fig_01},
 except the variations are much weaker
 (cf. the dramatically expanded scale)
 and the threshold behavior is different. Namely, the spectroscopic factor is continuous and  smooth;
 it is its derivative that exhibits a cusp  around the $0d_{3/2}$
 threshold. Again, this is consistent with the general expectation
 that, for $\ell$=2,  $Re(S^2)$ {\em below} the 1n threshold of $^{17}$O
 behaves as $(-e_{\ell j})^{3/2}$
 while {\em above} the threshold $Re(S^2)$ ($Im(S^2)$) should
 behave as $(e_{\ell j})^{3/2}$ ($(e_{\ell j})^{5/2}$).
 The associated 1n $d_{3/2}$  radial overlap integrals (not displayed)
 are extremely close to the s.p.~resonant $0d_{3/2}$ wave functions.

 \begin{figure}[htb]
 \centering
 \includegraphics*[width=10cm]{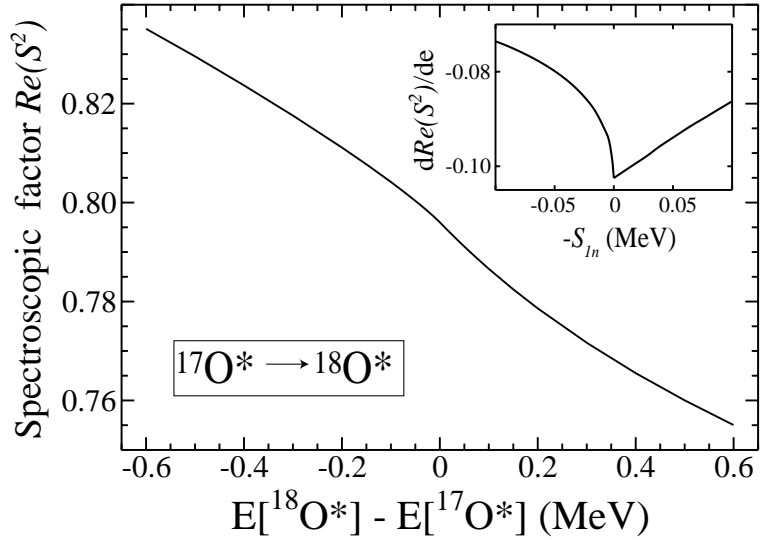}
 \caption{Similar to Fig.~\ref{fig_03} except for the overlap
 in the excited  $0^+_3$ state of $^{18}$O: $\displaystyle \langle
 ^{18}{\rm O}( 0^+_3) | [^{17}{\rm O}( 3/2_1^+) \otimes d_{3/2}]^{0^+}
 \rangle^2$. The energy derivative of the  spectroscopic factor around the 1n threshold
 is shown in the inset.}
 \label{fig_SF3}
 \end{figure}
 To investigate the behavior of spectroscopic factor in $^{18}$O around $S_{1n}$=0, the
 depth of the WS potential has been decreased to lower the position
 of the $0d_{3/2}$ s.p. pole down to --5 MeV. At the same time, the
 SDI coupling strength $V_0$ has been  modified  to allow $S_{1n}$ of
 the $0^+_3$ state of $^{18}$O  to go through zero. The results displayed in
 Fig.~\ref{fig_SF3} are consistent with the Wigner limit for
 $\ell$=2. Namely,  spectroscopic factor and its first derivative are continuous around
 $S_{1n}$=0, while the second derivative exhibits discontinuity (see the
 inset in Fig.~\ref{fig_04}). The associated 1n $d_{3/2}$  radial overlap 
 integrals (not displayed)
are extremely close to the s.p. resonant $0d_{3/2}$ wave function of a
real (or  complex) WS potential.

 \section{Conclusions} \label{conclusion}

By explicit many-body GSM calculations that fully account for a  
coupling to the scattering space,  we demonstrate the presence of a
near-threshold, non-perturbative rearrangement in the wave function that
has an appreciable  low angular momentum s.p. component. The threshold
behavior of spectroscopic factors (Wigner cusp) can only be reproduced 
through a complete inclusion of scattering states, including the
non-resonant space, in the GSM basis. Having a complete basis
which allows to describe bound, weakly-bound and unbound states on the
same footing, is the only way to guarantee the unitarity, which lies at
the basis of the Wigner threshold effect in cross sections and other
observables, in particular in the  multichannel case. This fundamental
requirement is not respected in any CQS formulation of the many-body
theory. As shown in several examples discussed in our work, restoration
of unitarity can strongly affect both values and behavior of  spectroscopic factors.

 The detailed analysis of near-threshold behavior of spectroscopic factors
associated with  $\ell$=1 (heliums) and $\ell$=2 (oxygens) partial waves
 shows the presence of the Wigner limit in the many-body GSM solution. Namely,
the fluctuating  component of the spectroscopic factor  behaves as $(-e_{\ell
j})^{\ell-1/2}$ 
below the 1n threshold.  Above the  threshold, the spectroscopic factor  is complex;
the real part behaves as $(e_{\ell j})^{\ell+1/2}$ while the imaginary
part, associated with the decaying nature of a resonance, varies as
$(e_{\ell j})^{\ell-1/2}$. 

If either parent or daughter nucleus is unbound, the corresponding
one-nucleon overlap integral is complex. It can be very well
approximated by a resonant state of a one-body potential which
reproduces the complex generalized one-nucleon  separation energy 
${\tilde S}_{1n}$ (\ref{genSn}). This is a straightforward generalization
of an approximate treatment of one-nucleon overlaps often used in SM
studies (see, e.g., Ref.~\cite{sf3}) where a real average potential is
employed with a depth adjusted to reproduce $S_{1n}$. Therefore, we
conclude  that the realistic radial overlap functions, which constitute
the basic theoretical input needed for a description of transfer 
reactions, can be conveniently generated by a radial s.p. wave function
of the (complex) one-body  potential reproducing the generalized 
separation energy. On the other hand, the normalization of the overlap
integral, i.e., the spectroscopic factor, cannot be obtained in a simple
way; here the full microscopic treatment is necessary (see discussion in
Ref.~\cite{Mic07}).

Optimally, the complex average potential used to generate one-nucleon
overlap functions should be generated self-consistently using the
Gamow-Hartree-Fock procedure \cite{Mic04}. The imaginary part of this
potential should not be confused with the absorption potential used in
the context of optical model studies. As we illustrated in this paper,
depending on $ {\tilde S}_{1n}$, the corresponding resonant states do
not always obey the usual Berggren classification for hermitian
Hamiltonians. For instance, one has to consider decaying bound states
(resonant states lying
in the first quarter of the complex-$k$ plane) or threshold 
states that have zero energy but nonzero width.

 While Berggren never considered complex potentials in the
derivation of completeness relation for Gamow states
\cite{Berggren1,Berggren2}, its demonstration in the case of localized 
potentials, as those of the WS  type used in this paper, is
straightforward~\cite{Niko73}. For  complex potentials, multiple
$S$-matrix poles can appear in the resonant-state expansions. This
complicates the completeness relation as some additional states, which
are not eigenfunctions of the one-body Hamiltonian, have to be added to
reach completeness \cite{Niko73}. Moreover, a possible appearance of the
spectral singularities, i.e., the $S$-matrix poles lying on the real
momentum axis, not considered in Ref.~\cite{Niko73}, can complicate 
matters \cite{Samsonov}. However, multiple singularities cannot happen
if the imaginary part of the potential is small enough, and - in this
case - the Berggren completeness relation for complex potentials is
obtained in the same manner as for real potentials. Such a 
relation could be useful when developing the  generalized Gamow-Hartree-Fock method 
that deals with unusual Gamow states.

\vskip 1truecm
This work was supported by  the U.S. Department of Energy
under Contracts Nos. DE-FG02-96ER40963 (University of Tennessee),
DE-AC05-00OR22725 with UT-Battelle, LLC (Oak Ridge National
Laboratory), and DE-FG05-87ER40361 (Joint Institute for Heavy Ion Research).


\begin{thebibliography}{99}
\bibitem{doba} J. Dobaczewski, and W. Nazarewicz,
     Phil. Trans. R. Soc. Lond. A 356 (1998) 2007.
\bibitem{brown}  B.A. Brown, Prog. Part. Nucl.  Phys. {\bf 47} (2001) 517.
\bibitem{opr} J. Oko{\l}owicz, M. P{\l}oszajczak, and I. Rotter, Phys. Rep. {\bf 374} (2003) 271.
\bibitem{cau} E. Caurier, G. Mart'nez-Pinedo, F. Nowacki, A. Poves,
and A.P. Zuker,  Rev. Mod. Phys. {\bf 77} (2005) 427.
\bibitem{te} R.G. Thomas, Phys. Rev. {\bf 81} (1951) 148; \\
R.G. Thomas, {\bf 88} (1952) 1109; \\
J.B. Ehrman, Phys. Rev. {\bf 81} (1951) 412.
\bibitem{ppnp} J. Dobaczewski, N. Michel, W. Nazarewicz, M. P{\l}oszajczak, and
J. Rotureau, Prog. Part. Nucl. Phys. {\bf 59} (2007) 432.
\bibitem{Mic02} N. Michel, W. Nazarewicz, M. P{\l}oszajczak, and K. Bennaceur, Phys. Rev. Lett. {\bf 89} (2002) 042502; \\
N. Michel, W. Nazarewicz, M. P{\l}oszajczak, and J. Oko{\l}owicz, Phys. Rev. C {\bf 67} (2003) 054311.
\bibitem{Mic04}
N. Michel, W. Nazarewicz, and M. P{\l}oszajczak, Phys. Rev. C {\bf 70} (2004) 064313.
\bibitem{Bet02}  R.I. Betan, R.J. Liotta, N. Sandulescu, and T. Vertse, Phys. Rev. Lett. {\bf 89} (2002) 042501; \\
 R.I. Betan, R.J. Liotta, N. Sandulescu, and T. Vertse, Phys. Rev. C {\bf 67} (2003) 014322.
 \bibitem{Mic07} N. Michel, W. Nazarewicz, and M. P{\l}oszajczak, 
Phys. Rev.  C {\bf 75} (2007) 031301(R).
\bibitem{Berggren1} T. Berggren, Nucl. Phys. A {\bf 109} (1968) 265.
\bibitem{Berggren2} T. Berggren and P. Lind, Phys. Rev. C {\bf 47} (1993) 768.
\bibitem{Gel64} I. Gelfand and N. Ya. Vilenkin, {\it Generalized Functions}, Volume 4, (Academic Press, New York 1964).
\bibitem{Bohm} A. Bohm, {\it The Rigged Hilbert Space and Quantum Mechanics},
 Lecture Notes in Physics {\bf 78} (Springer, New York  1978).
\bibitem{Madrid} R. de la Madrid, Eur. J. Phys. {\bf 26} (2005) 287.
\bibitem{CivGad} O. Civitarese and M. Gadella, Phys. Rep. {\bf 396} (2004) 41.
\bibitem{HagenVaagen} G. Hagen and J. Vaagen, Phys. Rev. C {\bf 73} (2006) 034321.
\bibitem{diagnumbook} R. Freund, Lanczos Method for Complex Symmetric Eigenproblems (Section 7.11),
in Z. Bai, J. Demmel, J. Dongarra, A. Ruhe, and H. van der Vorst, editors, {\it Templates for the Solution of Algebraic Eigenvalue Problems: A Practical Guide},
SIAM, Philadelphia, 2000.
\bibitem{dmrg1} S.R. White, Phys. Rev. Lett. {\bf 69} (1992) 2363; 
S.R. White, Phys. Rev. B  {\bf 48} (1993) 10345.
\bibitem{dmrg2} J. Rotureau, N. Michel, W. Nazarewicz, M. P{\l}oszajczak, and J. Dukelsky,  Phys. Rev. Lett. {\bf 97} (2006) 110603.
\bibitem{Civ99} O. Civitarese, M. Gadella, and R. Id Betan,
    Nucl. Phys. A {\bf 660} (1999) 255.
\bibitem{satch} G.R. Satchler, {\it Direct Nuclear Reactions}
(Clarendon Press, Oxford, 1983).
\bibitem{sfs} M.H. Macfarlane and J.B. French, Rev. Mod. Phys. {\bf 32} (1960) 567.
\bibitem{Glenden} N.K. Glendenning, Ann. Rev. Nucl. Sci. {\bf 13} (1963) 191; \\
         N.K. Glendenning, {\it Direct Nuclear Reactions} (Academic Press Inc., 1983).
\bibitem{FroLipp} P. Fr{\"o}brich and R. Lipperheide, {\it Theory of Nuclear Reactions} (Oxford Science Publications, Clarendon Press Oxford, 1996)
\bibitem{sf1} Jenny Lee {\it et al.}, arXiv:nucl-ex/0511023;\\
M.B. Tsang, Jenny Lee, and W. G. Lynch, Phys. Rev. Lett. {\bf 95} (2005) 222501.
\bibitem{sf2}  P.G. Hansen and J.A. Tostevin, Ann. Rev. Nucl. Part. Sci. {\bf 53} (2003) 219.
\bibitem{sf3} A. Gade {\it et al.},  Eur. Phys. J. A {\bf 25} (2005) s01, 251;\\
D. Bazin {\it et al.}, Phys. Rev. Lett. {\bf 91} (2003) 012501.
\bibitem{model_indep_exp} A.M. Mukhamedzhanov and F.M. Nunes, Phys. Rev. C {\bf 72} (2005) 017602.
\bibitem{Fur02} R.J. Furnstahl and  H.-W. Hammer, Phys.Lett. B {\bf 531} (2002) 203.
\bibitem{modeldep} J.A. Tostevin, J. Phys. G: Nucl. Part. Phys. {\bf 25} (1999) 735; \\
V. Maddalena {\it et al.}, Phys. Rev. C {\bf 63} (2001) 024613; \\
B.A. Brown, P.G. Hansen, B.M. Sherrill, and J.A. Tostevin, Phys. Rev. C {\bf 65} (2002) 061601.
\bibitem{Bohr} A. Bohr and B.R. Mottelson, {\it Nuclear Structure}, Vol. 1 (New York, W.A. Benjamin, 1969).
\bibitem{Gyarmati} B. Gyarmati, F. Krisztinkovics and T. Vertse, Phys. Lett. B {\bf 41} (1972) 110.
\bibitem{Berggren3} T. Berggren, Phys. Lett. B {\bf 373} (1996) 1.
\bibitem{Wigner} E.P. Wigner, Phys. Rev. {\bf 73} (1948) 1002.
\bibitem{Hategan1} C.~Hategan, Annals of Physics  {\bf 116} (1978) 77.
\bibitem{Graw} G. Graw and C.~Hategan, Phys. Lett. B {\bf 37} (1971) 41.
\bibitem{Op05} J. Oko{\l}owicz and M. P{\l}oszajczak,  Int. J. Mod. Phys. E {\bf 15} (2006) 529; \\
Y. Luo, J. Okolowicz, M. Ploszajczak, and N. Michel, arXiv:nucl-th/0211068.
\bibitem{Breit} G. Breit, Phys. Rev. {\bf 107} (1957) 1612.
\bibitem{Baz} A.I. Baz, Soviet Phys. - JETP {\bf 6} (1957) 709.
\bibitem{Newton} R.G. Newton, Phys. Rev. {\bf 114} (1959) 1611.
\bibitem{Meyerhof}  W.E. Meyerhof, Phys. Rev. {\bf 129} (1963) 692.
\bibitem{Baz1}  A.I.~Baz, Y.B.~Zeldovich,  and A.M.~Peremolov,
 {\it Scattering, Reactions and Decays in Nonrelativistic Quantum Mechanics},
(Jerusalem, 1969, translated from Russian, Nauka, Moscow, 1966).
\bibitem{Lane}  A.M. Lane, Phys. Lett. B {\bf 33} (1970) 274.
\bibitem{Hategan} C. Hategan, Proc. Rom. Acad. A {\bf 3} (2002) 11.
\bibitem{Malmberg} P.R. Malmberg, Phys. Rev. {\bf 101} (1956) 114.
\bibitem{Wells} J.T. Wells, A.B. Tucker, and W.E. Meyerhof, Phys. Rev. {\bf 131} (1963) 1644.
\bibitem{Moore} C.F. Moore {\it et al.},  Phys. Rev. Lett. {\bf 17} (1966) 926.
\bibitem{Abramovich} S. Abramovich, B. Guzhovskij, and L. Lasarev, Sov. J. Part. Nucl. {\bf 23} (1992) 129.
\bibitem{Dic04} W.H. Dickhoff and C. Barbieri, Prog. Part. Nucl. Phys. {\bf 52} (2004) 377.
\bibitem{Blo77} L.D. Blokhintsev, I. Borbely, and E.I. Dolinskii, Sov. J. Part.
  Nucl. {\bf 8}  (1977)c485.
\bibitem{Tim03} N.K. Timofeyuk, L.D. Blokhintsev, and J.A. Tostevin, Phys. Rev.
  C {\bf 68} (2003) 021601(R).
 \bibitem{Bay96}  D. Baye, G. Levai, and J.-M. Sparenberg, 
  Nucl. Phys. A {\bf 599} (1996) 435.
\bibitem{Cha01} V.M. Chabanov and B.N. Zakhariev, Inverse Problems 17 (2001) 683. 
 \bibitem{Niko73}  M.V.~Nikolayev and  V.S.~Olkhovsky, Lett. Nuo. Cime. {\bf 8}
  (1973) 703.
\bibitem{Samsonov} B.F.~Samsonov, J. Phys. A: Math. Gen. {\bf 38} (2005) L571.

\end{thebibliography}
\end{document}